\documentclass[aps,prl,twocolumn,superscriptaddress,10pt,reprint]{revtex4-1}

\newcommand{\ket}[1]{\ensuremath{|#1\rangle}}
\newcommand{\bra}[1]{\ensuremath{\langle #1 |}}
\newcommand{\vro}{\hat{\varrho}}

\usepackage[pdftex]{graphicx}

\newcommand{\bc}{\begin{center}}
\newcommand{\ec}{\end{center}}

\begin{document}
\title{Single spontaneous photon as a coherent beamsplitter for an atomic matterwave}

\author{Ji\v r\'{i} Tomkovi\v c}
\email[E-Mail: ]{single-photon@matterwave.de}
\affiliation{Kirchhoff-Institut f\"{u}r Physik, Universit\"{a}t Heidelberg, Im Neuenheimer Feld 227, 69120 Heidelberg, Germany}
\author{Michael Schreiber}
\affiliation{Ludwig-Maximilians-Universit\"{a}t, Schellingstr. 4, 80799 M\"{u}nchen, Germany}
\author{Joachim Welte}
\affiliation{Kirchhoff-Institut f\"{u}r Physik, Universit\"{a}t Heidelberg, Im Neuenheimer Feld 227, 69120 Heidelberg, Germany}
\author{Martin Kiffner}
\affiliation{Physik Department I, Technische Universit\"{a}t M\"{u}nchen,\\ James-Franck-Stra{\ss}e, 85747 Garching, Germany}
\author{J\"{o}rg Schmiedmayer}
\affiliation{Vienna Center for Quantum Science and Technology, Atominstitut, TU Wien, 1020 Vienna, Austria}
\author{Markus K.\ Oberthaler}
\affiliation{Kirchhoff-Institut f\"{u}r Physik, Universit\"{a}t Heidelberg, Im Neuenheimer Feld 227, 69120 Heidelberg, Germany}

\begin{abstract}
    In spontaneous emission an atom in an excited state undergoes a transition to the ground state and emits a single photon. Associated with the emission is a change of the atomic momentum due to photon recoil \cite{SPBS:Milonni_Book}. Photon emission can be modified close to surfaces \cite{SPBS:Milonni_OpticsComm73,SPBS:Drexhage_1974} and in cavities \cite{SPBS:Goy1983}. For an ion, localized in front of a mirror, coherence of the emitted resonance fluorescence has been reported \cite{SPBS:Eschner_Nature2001,SPBS:Eschner_EPJD2003}. In free space experiments demonstrated that spontaneous emission destroys motional coherence \cite{PhysRevLett.73.1223,SPBS:Schmiedmayer_Prl1995,PhysRevLett.86.2191}. Here we report on motional coherence created by a single spontaneous emission event close to a mirror surface. The coherence in the free atomic motion is verified by atom interferometry \cite{SPBS:Oberthaler_1999}. The photon can be regarded as a beamsplitter for an atomic matterwave and consequently our experiment extends the original recoiling slit Gedanken experiment by Einstein \cite{SPBS:Bohr_1949,Bertet2001} to the case where the slit is in a robust coherent superposition of the two recoils associated with the two paths of the quanta.
\end{abstract}

\maketitle

\begin{figure*}[!bt]
     \includegraphics[scale=1]{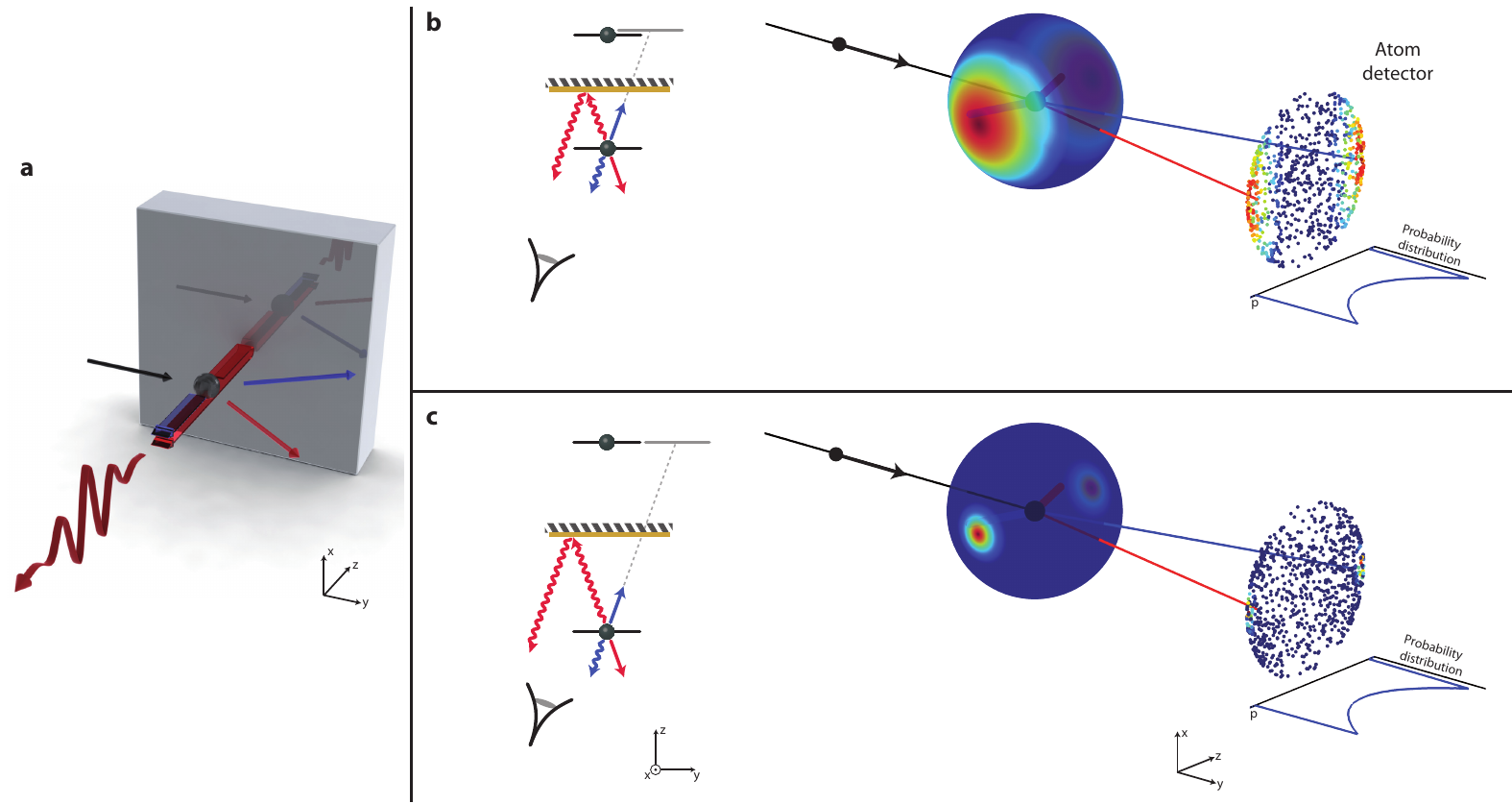}
\caption{Motional coherence generated by a single spontaneous emission event. (a) The situation of interest is depicted -- an atom in front of a mirror spontaneously emits a single photon. For emission perpendicular to the mirror surface an observer can in principle not distinguish if the photon has been reflected or not. Momentum conservation in the atom-photon system implies that the atom after the emission  is in a coherent superposition of two different momentum states separated by twice the photon recoil. (b) Indistinguishability is also given for more general emission directions.
With the spatial extension of the atom corresponding to the optical absorption cross section, indistinguishability can be estimated by the projected overlap of atom and its mirror-image.
This overlap is represented colorcoded on a sphere for all emission directions (red: full coherence, blue: no coherence). Repeating the experiment -- single atom emits a single photon -- leads to the indicated pattern at the atom detector. The colorcode indicates the probability generating a coherent superposition for the corresponding event (red: full coherence, blue: no coherence). (c) In the case of large distances to the mirror the coherent portion drastically reduces, approaching the limit of vanishing coherence in free space.}
\end{figure*}

We consider an atom passing by a mirror which spontaneously emits a single photon (see Fig.\ 1a). Due to the photon momentum the atom gets a corresponding recoil kick  in the direction opposite to the photon emission. In the absence of the mirror the observation of the emitted photon direction implies the knowledge of the atomic momentum resulting from the photon-atom entanglement \cite{SPBS:Schmiedmayer_Prl1995}. In the presence of the mirror the detection of a photon in a certain direction does not necessarily reveal if it has reached the observer directly or via the mirror. For the special case of spontaneous emission perpendicular to the mirror surface the two emission paths are in principle not distinguishable for small atom-mirror distances $d \ll c/\Gamma$  with $c$ the speed of light and $\Gamma$ the natural linewidth. This general limit is always fulfilled in our experiments. Thus the atom after this emission event is in a superposition of two motional states.
\begin{figure*}[!t]
    \includegraphics[scale=1]{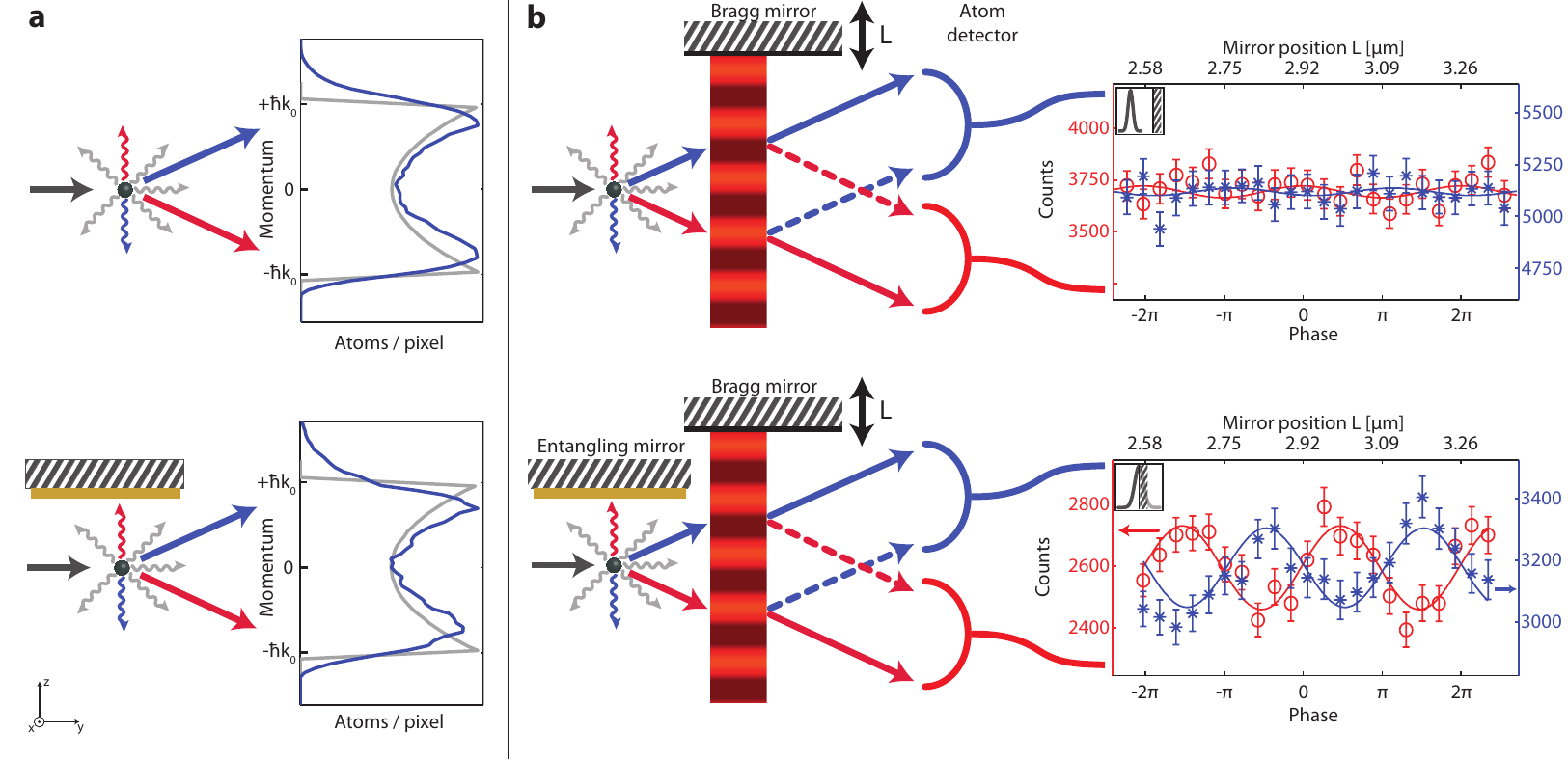}
    \caption{Experimental confirmation of coherence induced by spontaneous emission. (a) Experimental observation of momentum distribution does not reveal the coherence. In both cases - close to and far from the mirror - the momentum distribution is the same (blue line). In order to compare the observed momentum distribution after spontaneous emission with theory (light gray) the data has been deconvoluted by the initial momentum distribution. The deviation results from  a residual filtering of high spatial frequencies. (b) The coherence is revealed if the spontaneous emission event is employed as the first beamsplitter of an atom interferometer. The recombination is accomplished by Bragg scattering from a standing light wave. The relative phase of the two paths can be changed by moving the ''Bragg'' mirror as indicated. In the case of a mean distance of $54 \ \mu m$ between atoms and ''entangling'' mirror (upper graph, error bars indicate poisson noise) no interference signal is observed confirming the free space limit.
    The inset depicts the position of ''entangling'' mirror to the atomic beam.
    For a mean distance of $2.8 \ \mu m$ (lower graph) the two complementary outputs of the interferometer reveal an interference pattern with a maximal visibility of $5.9\% \pm 1.1 \% $.}
\end{figure*}

This is also true for the more general case of tilted emission as revealed in Fig.\ 1b for emission close to the mirror surface. One expects residual coherence for emission angles where the optical absorption cross section of the atom and the mirror-atom observed by a fictitious observer in the emission direction still overlap. This is visualized in Fig. 1b, where the corresponding cross sections are indicated with the bars. The overlap as a function of emission direction is depicted on the sphere (blue no coherence, red full coherence). The result on the atomic motion is indicated for one special trajectory which starts with an atom moving parallel to the mirror surface and a single photon emission under an angle to the mirror normal. This case leads to an imperfect coherent superposition of two momentum states separated by less than two photon momenta $\hbar k_0$. The spatial distribution of the atoms at the position of the detector is shown, where the color corresponds to the degree of coherence. In Fig.\ 1c we contrast this to the case of larger distance to the mirror, where the portion of coherent atomic momentum is strongly reduced.

It is important to keep in mind that a single particle detector cannot distinguish between coherent superpositions and mixtures but only gives the probability distribution. Thus an interferometric measurement \cite{SPBS:Cronin_2009} has to be applied to reveal the expected coherent structure (see Fig.\ 2). For that, the two momentum states of interest have to be overlapped and the coherence i.e.\ well defined phase difference, is verified by observing an interference pattern as function of a controlled phase shift applied to one of the momentum states.
The two outermost momentum states are expected to show the highest coherence. Their recombination can be achieved by a subsequent Bragg scattering off an independent standing light wave (see Fig. 2b) with the suitable wavelength
\cite{SPBS:Oberthaler_1999, SPBS:Martin_1988}.
The relative phase $\phi_{\mathrm{B}}$ is straightforwardly changed shifting the probing standing light wave. This is implemented by moving the retroreflecting mirror by distance $L$. The upper graph depicts the results obtained for large distances \mbox{($>54 \mu m$)} of the atom to the mirror i.e.\ a free atom. In this case no interference is observed, and thus spontaneous emission induces a fully incoherent modification of the atomic motion. For a mean distance of $2.8 \mu m$ clear interference fringes are observed demonstrating that a single spontaneous emission event close to a mirror leads to a coherent superposition of outgoing momentum states.

In the following we describe the essential parts of experimental setup shown in Fig.\ 2b, lower graph. Further details are provided in the supplementary information. Since the effect critically depends on the distance between atom and mirror a well collimated and localized beam of  $^{40}Ar$ atoms in the metastable $1s_5$ state is used. In order to ensure the emission of only a single photon we induce a transition $1s_5 \rightarrow 2p_4$ ($\lambda_E = 715 nm$). From the excited state $2p_4$ the atom predominantly decays to the metastable $1s_3$ state via spontaneous emission of a single photon ($\lambda_{SE} = 795 nm$) (branching ratio of $1s_5/1s_3 = 1/30$). The residual $1s_5$ are quenched to an undetectable ground state with an additional laser. Choosing the appropriate polarization of the excitation laser the atomic dipole moment is aligned within the mirror plane leading to the momentum distribution after spontaneous emission shown in Fig.\ 2a. The interferometer is realized with a far detuned standing light wave on a second mirror. Finally the momentum distribution is detected by a spatially resolved multi channel plate (MCP) approx.\ $1m$ behind the spontaneous emission  enabling to distinguish between different momenta.

For  systematic studies of the coherence we analyze the probability for finding a particle in a coherent superposition of momentum states as a function of atom-mirror distance $d$. This is done by analyzing the final momentum distribution for different phases $\phi_{\mathrm{B}}$ within the interferometer and fit for each resolved momentum ($\approx 1/8 $ of a photon momentum) an interference pattern given by
\begin{equation}
 N = N_0 + N_A \cos(\phi_{\mathrm{B}} +\phi_0) .
\end{equation}
In Fig.\ 3 we plot the visibility $V=N_A/N_0$ (with $N_0$ the constant atom number, $N_A$ the oscillatory part) revealing that the coherence vanishes within distances of a few micrometers to the mirror.

For a basic understanding of the physics behind the experimental observation we use a simple semiclassical model. We follow the picture of an atom and its image by Morawitz \cite{SPBS:Morawitz_PR1969} and Milonni, Knight \cite{SPBS:Milonni_OpticsComm73} and assume a two level system with ground state $\ket{g}$ and excited $\ket{e}$. In order to deduce the indistinguishability between the atom and its mirror atom, i.e.\ the photon emission towards and away from the mirror, we attribute to the atom a size corresponding the optical absorption cross section ($\sigma = 3 \lambda^2 / 2 \pi$). In the direction perpendicular to the mirror an observer can not distinguish atom and mirror atom in principle and thus a coherent superposition of momentum states is emerging $\ket{\psi}=1/\sqrt{2}(\ket{+\hbar k_0} + \ket{- \hbar k_0})$ with the photon momentum $p_{rec}=\hbar k_0$.
For emission directions other than perpendicular the probability $P$ for generating  $\ket{\psi'}=1/\sqrt{2}(\ket{+\hbar k'} + \ket{- \hbar k'})$ can be estimated by the overlap region of atom and mirror atom with the assigned effective size as shown in Fig.\ 1b. This overlap depends on the distance between atom and mirror and on the observation angle  (for details see supplementary information).
In order to quantitatively compare with the experimental data the finite resolution of momentum detection has to be taken into account leading to an integration over different observation directions. Further averaging due to the finite extension of the atomic beam (width in transverse direction of $10\mu m$) and the initial momentum distribution results in a reduction of the visibility $V$. The prediction within this model is shown as solid blue line in Fig.\ 3.
\begin{figure}[!h]
    \includegraphics[scale=1]{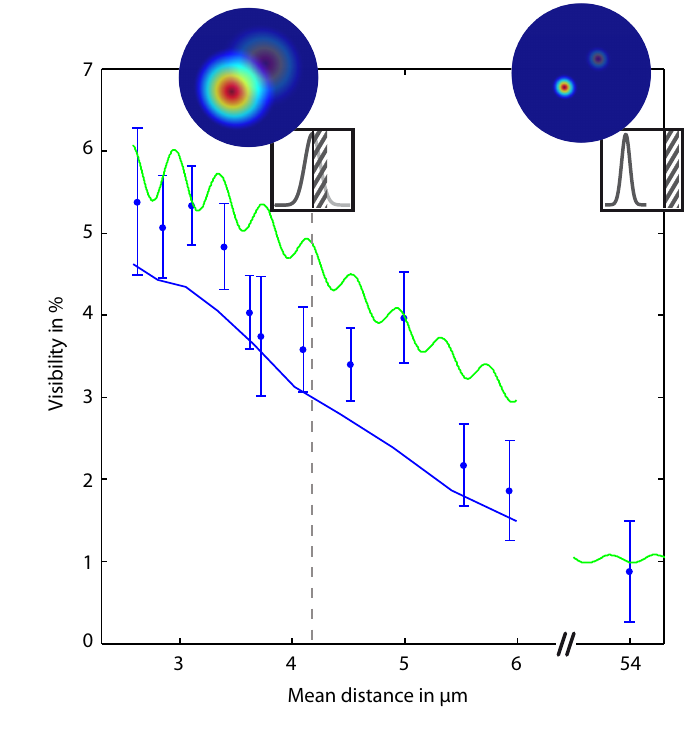}
    \caption{Dependence of visibility on the mean atom-mirror distance.  Measured data is depicted as blue points.
    The mean distance is calculated from the position of the ''entangling'' mirror with respect to the center of the atomic beam as indicated in the insets.    The error bars indicate a $95\%$ confidential interval resulting from the fitting procedure to the interference pattern. The expectation from the simple cross section overlap model is shown with the blue line. The quantum mechanical treatment is depicted as green line. One finds good agreement between theory and experiment by including details such as initial spatial and momentum distribution, averaging over all distances, details of Bragg scattering and the final spatial resolution of the atom detector. The mean atom-mirror distance is adjusted by the position of the ''entangling'' mirror with respect to the collimation slit of the atomic beam.}
\end{figure}

The comprehensive quantum mechanical model (for details see supplementary information) takes into account the modified mode structure of the electromagnetic field due to the presence of the mirror \cite{SPBS:stefano:00}. We derive a master equation for the internal degrees of freedom of the atom and its center of mass motion perpendicular to the mirror surface. It is found that the quantum state of the atomic center of mass motion after spontaneous emission can  be written as
\begin{equation}
    \vro_{gg}(t=\infty) =  \alpha \frac{3}{8} \int\limits_0^1 \mathrm{d}u
    \left( \ket{\psi_s}\bra{\psi_s} +u^2 \ket{\psi_p}\bra{\psi_p} \right), \label{rhopara2}
\end{equation}
where
\begin{eqnarray}
    \ket{\psi_s} &=& \left(r_s^* e^{ik_0 u\hat{z}} + e^{-ik_0 u \hat{z}} \right)\ket{\psi_0}, \label{state1}\\
    \ket{\psi_p} &=& \left(-r_p^* e^{ik_0 u\hat{z}} + e^{-ik_0 u \hat{z}}\right)\ket{\psi_0}. \label{state2}
\end{eqnarray}
The operators $e^{\pm ik_0 u \hat{z}}$ in Eqs.~(\ref{state1}) and~(\ref{state2}) describe the
transverse recoil momentum $\pm \hbar k_0 u$ transferred to the atom by the spontaneously emitted photon. The Fresnel coefficient $r_s$ ($r_p$) accounts for the reflection of the transversal electric (transversal magnetic) mode at the mirror and $\ket{\psi_0}$ describes the motional state of the atom before spontaneous emission. The normalization is ensured by the normalization constant $\alpha$.
For a quantitative comparison with the experiment  we assume that
\begin{equation}
\ket{\psi_0} = \int \mathrm{d} p f(p,d) e^{\frac{i}{\hbar}p d}e^{i\phi_f(p)}\ket{p}
\label{psi0}
\end{equation}
is a coherent wave packet. The quantity $|f(p,d)|^2$ represents the initial momentum distribution of atoms and is inferred from an independent measurement of the momentum distribution. The description of the initial atomic state by a pure state is a sensible assumption since the width of the slit collimating the atoms is chosen to be close to the diffraction limit. The phase $\phi_f(p)$ determines the shape of the wavefunction in position space.
The Bragg grating is modeled as a beamsplitter with a momentum dependent splitting ratio determined from experimental measurements. After free evolution of the atom we determine the probability to detect the atom within the given resolution of the detector. The result of this calculation is shown as green line in Fig.\ 3
where only the phase $\phi_f(p)$ of the wavefunction in front of the first mirror cannot be fully reconstructed acting as a free parameter. The uncertainty of this phase explains a smaller visibility and the asymmetry between different diffraction orders (see Fig.\ 4).

So far we have discussed the maximum coherence observed in the experiment. In Fig.\ 4 the momentum dependence of the coherence is shown for a mean distance of $3.3 \mu m $ from the mirror.
This reveals that only the outermost parts of the momentum distribution are in a coherent superposition which is consistent with the simple picture of atom and mirror atom. It is important to note that Bragg scattering itself exhibits a momentum dependence (Bragg acceptance). For the chosen short interaction length the Bragg acceptance is indicated by the gray line in Fig.\ 4. Since the observed coherence decays significantly within the Bragg acceptance we can experimentally confirm that only the most extreme emission events i.e.\ perpendicular to the mirror surface, lead to a significant generation of coherence. This angular dependence is similar for all investigated mirror distances since it is essentially given by the coherent momentum spread of the strongly confined initial atomic beam.
\begin{figure}[!h]
    \includegraphics[scale=1]{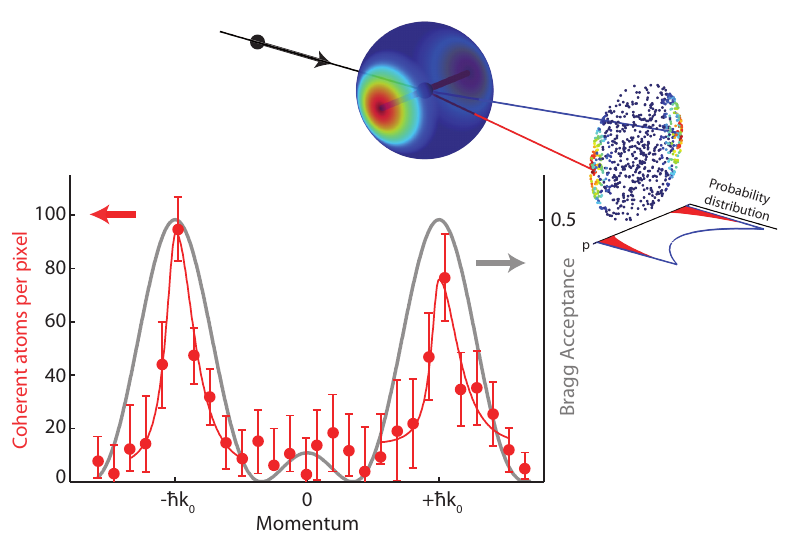}
    \caption{Observation of angular dependence of coherence. The schematics show an idealized case of coherent momenta for an atom in a fixed distance and an initial momentum parallel to the mirror (red area within the momentum distribution). Due to finite momentum distribution of the atomic beam, the narrow coherent momenta is smeared out in the experimental realization. The measured width of coherent momenta (red points) is smaller than the angle-acceptance of the Bragg-crystal (gray line), revealing that mainly atoms with momenta of $\pm \hbar k_0$ are in a coherent superposition. The data is shown for a mean distance of $3.3 \  \mu m $ (in contrast to Fig.\ 2b (lower graph),  where the atom is much closer to the mirror). Error bars are defined accordingly to Fig. 3.}
\end{figure}

Finally we would like to point out the differences to other experiments where the connection between spontaneous emission and coherence is investigated. For example the experiment in \cite{SPBS:Schmiedmayer_Prl1995} shows that the
spontaneous photon carries away information from the atom about its position, and therefore destroys the coherence when the two paths can be distinguished.
The experiment \cite{SPBS:Eschner_Nature2001} on the other hand provides direct proof for the coherence of photons emitted in the  resonance fluorescence of a laser-driven ion in front of a mirror. The observed interference pattern can be regarded as indirect evidence for  the  motional coherence of the trapped ion, well within the Lamb-Dicke limit \cite{SPBS:Eschner_EPJD2003}.
A different example in the context of laser cooling is velocity selective coherent population trapping \cite{SPBS:vscpt} where spontaneous emission populates motional dark states. Here the direction of the emitted photon is indistinguishable since it is emitted in the direction of a macroscopic classical field that drives the atom.
The  most salient feature of  our experiment is that a single spontaneous emission event in front of a mirror creates a coherent superposition in freely propagating atomic matter waves, without any external coherent fields involved. The emission directions of the spontaneous photon become indistinguishable due to the mirror.

In the work by Bertet et al. \cite{Bertet2001} photons from transitions between internal states are emitted into a high finesse cavity. Their first experiment reported in \cite{Bertet2001} demonstrates the transition from indistinguishability when emission is into a large \emph{classical} field to distinguishability and destruction of coherence between the internal atomic states when emission is into the vacuum state of the cavity. In their second experiment \cite{Bertet2001} they show that, using the same photon for both beamsplitters in an internal state interferometer sequence, coherence can be obtained even in the empty cavity limit. In our experiment the photon leaves the apparatus. We observe coherence only when the photon cannot carry away which-path information. This implies that the generated coherence in motional states is robust and lasts. In this sense it is an extension of Einstein's famous recoiling slit Ge\-danken experiment \cite{SPBS:Bohr_1949}. The single photon is the ultimate light weight beamsplitter which can be in a robust coherent superposition of two motional states. In free space the momentum of the emitted photon allows to measure the path of the atom. This corresponds to a well defined motional state of the beamsplitter i.e. no coherence. Close to the mirror the reflection renders some paths indistinguishable realizing a coherent superposition of the beamsplitter. The large mass of the mirror ensures that even in principle the photon recoil cannot be seen. Thus the atom is in a coherent superposition of the two paths. We measure this generated coherence by matterwave interference.


\begin{acknowledgments}
    We wish to thank Florian Ritterbusch for assistance throughout the preparation of this manuscript.

    We gratefully acknowledge support from the Forschergruppe FOR760, Deut\-sche Forschungsgemeinschaft, the German-Israeli Foundation, the Heidelberg Center of Quantum Dynamics, Landesstiftung Baden-W\"{u}rttem\-berg, the ExtreMe Matter Institute and the European Commission Future and Emerging Technologies Open Scheme project MIDAS (Macroscopic Interference Devices for Atomic and Solid-State Systems).

    M.\ K.\  acknowledges financial support within the framework of the Emmy Noether project HA 5593/1-1 funded by the German Research Foundation (DFG).

    J. S. acknowledges financial support through the Wittgenstein Prize.
\end{acknowledgments}


\end{document}